\documentclass[a4paper,11pt]{article}

\usepackage{jheppub} 
\usepackage{graphicx}
\usepackage{amsmath,amssymb,amsthm,bm,bigdelim,multirow}
\usepackage{color}
\usepackage{booktabs}
\usepackage{hyperref}
\usepackage{dsfont}
\usepackage{here}
\usepackage{subcaption}

\newcommand{\kslash}{k\kern-1ex /}
\newcommand{\pslash}{p\kern-1ex /}
\newcommand{\qslash}{q\kern-1ex /}
\newcommand{\lslash}{l\kern-1ex /}
\newcommand{\sslash}{s\kern-1ex /}
\newcommand{\Dslash}{D\kern-1.2ex /}

\newcommand{\beqa}{\begin{eqnarray}}
\newcommand{\eeqa}{\end{eqnarray}}
\newcommand{\Tr}{{\rm Tr}}

\newcommand{\bd}{\begin{description}}
\newcommand{\ed}{\end{description}}

\newcommand{\ben}{\begin{eqnarray}}
\newcommand{\een}{\end{eqnarray}}

\def\lsim{\raise0.3ex\hbox{$<$\kern-0.75em\raise-1.1ex\hbox{$\sim$}}}
\def\gsim{\raise0.3ex\hbox{$>$\kern-0.75em\raise-1.1ex\hbox{$\sim$}}}
\def\simgt{\rlap{\lower 3.5 pt\hbox{$\mathchar \sim$}}\raise 2.0pt \hbox {$>$}}
\def\simlt{\rlap{\lower 3.5 pt\hbox{$\mathchar \sim$}}\raise 2.0pt \hbox {$<$}}

\newcommand{\crr}{\color{red}}

\begin{document}

\title{Entanglement and R\'enyi entropies of (1+1)-dimensional O(3) nonlinear sigma model with tensor renormalization group}

\author[a]{Xiao Luo,}
	\affiliation[a]{Graduate School of Pure and Applied Sciences, University of Tsukuba, Tsukuba, Ibaraki
    305-8571, Japan}
    	\emailAdd{luo@het.ph.tsukuba.ac.jp}

  	\author[b]{Yoshinobu Kuramashi}
  	\affiliation[b]{Center for Computational Sciences, University of Tsukuba, Tsukuba, Ibaraki
    305-8577, Japan}
  	\emailAdd{kuramasi@het.ph.tsukuba.ac.jp}

        \abstract{
 We investigate the entanglement and R\'enyi entropies for the (1+1)-dimensional O(3) nonlinear sigma model using the tensor renormalization group method. The central charge is determined from the asymptotic scaling properties of both entropies. We also examine the consistency between the entanglement entropy and the $n$th-order R\'enyi entropy with $n\rightarrow 1$. 
}
\date{\today}

\preprint{UTHEP-781, UTCCS-P-148}

\maketitle

\section{Introduction}
\label{sec:intro}

In the past decade the tensor renormalization group (TRG) method \footnote{In this paper, the ``TRG method" or the ``TRG approach" refers to not only the original numerical algorithm proposed by Levin and Nave \cite{Levin:2006jai} but also its extensions \cite{PhysRevB.86.045139,Shimizu:2014uva,Sakai:2017jwp,Adachi:2019paf,Kadoh:2019kqk,Akiyama:2020soe,PhysRevB.105.L060402,Akiyama:2022pse}.}, which was originally proposed in the condensed matter physics~\cite{Levin:2006jai}, has been getting applied to the particle physics. Although the target models in the initial stage were restricted to the 2$d$ ones, recent studies cover various four-dimensional (4$d$) models with the scalar, gauge and fermion fields~\cite{Akiyama:2019xzy,Akiyama:2020ntf,Akiyama:2021zhf,Akiyama:2020soe,Akiyama:2022eip,Akiyama:2023hvt}. So far much attention has been paid to the sign-problem-free nature of the TRG method~\cite{Shimizu:2014uva,Shimizu:2014fsa,Kawauchi:2016xng,Kawauchi:2016dcg,Yang:2015rra,Shimizu:2017onf,Takeda:2014vwa,Kadoh:2018hqq,Kadoh:2019ube,Kuramashi:2019cgs,Nakayama:2021iyp}. On the other hand, there are few studies focusing on an ability of the direct evaluation of the partition function or the path-integral itself, which potentially allows us to measure the entanglement entropy ($S_A$) and $n$th-order R\'enyi entropy ($S_A^{(n)}$). Up to know only 2$d$ Ising and XY models were investigated~\cite{Ueda_2014,Yang:2015rra,Bazavov:2017hzi}. Note that it is difficult for the Monte Carlo method to measure the entanglement and R\'enyi entropies so that recent lattice QCD studies focused on the so-called entropic $C$-function, which is an UV-finite observable in proportion to $\partial_L S^{(2)}_A(L)$ with $L$ an interval of length, avoiding the direct measurement of the entanglement and R\'enyi entropies~\cite{Buividovich:2008kq,Itou:2015cyu,Rabenstein:2018bri}.

In this paper we measure the entanglement and R\'enyi entropies of the (1+1)$d$ O(3) nonlinear sigma model (O(3) NLSM) using the density matrix without resort to the transfer matrix formalism employed in Ref.~\cite{Bazavov:2017hzi}. This model is massive and shares the property of asymptotic freedom with the (3+1)$d$ non-Abelian gauge theories. We extarct the central charge from the entanglement and $n$th-order R\'enyi entropies using the scaling formula for the non-critical (1+1)$d$ models~\cite{Calabrese:2004eu}. The value of the central charge is comapred with the previous result obtained from the entanglement entropy with the matrix product state (MPS) method~\cite{Bruckmann:2018usp}. We also make a consistency check between the entanglement and R\'enyi entropies by extrapolating the $n$th-order R\'enyi entropy to $n=1$.

This paper is organized as follows. In Sec.~\ref{sec:method}, we define the (1+1)$d$ O(3) NLSM on the lattice and give the tensor network representation. We present the numerical results for the entanglement and R\'enyi entropies in Sec.~\ref{sec:results}. We determine the central charge and discuss the consistency  between the entanglement and R\'enyi entropies.
Section~\ref{sec:summary} is devoted to summary and outlook.

\section{Formulation and numerical algorithm}
\label{sec:method}

Although the definition of the (1+1)$d$ O(3) NLSM and its tensor network representation are already given in the appendix of Ref.~\cite{Luo:2022eje}, we briefly give the relevant expressions for this work to make this paper self-contained.   

\subsection{(1+1)-dimensional O(3) nonlinear sigma model}
\label{subsec:action}

We consider the partition function of the O(3) NLSM on an isotropic hypercubic lattice $\Lambda_{1+1}=\{(n_1,n_2)\ \vert n_1=1,\dots,2L, n_2=1,\dots,N_t\}$ whose volume is $V=(2L)\times N_t$. The lattice spacing $a$ is set to $a=1$ unless necessary. A real three-component unit vector $\bm{s}(n)$ resides on the sites $n$ and satisfies the periodic boundary conditions $\bm{s}(n+{\hat \nu}L)=\bm{s}(n)$ ($\nu=1,2$). The lattice action $S$ is defined as
\ben
\label{eq:action}
S&=&-\beta\sum_{n\in\Lambda_{1+1},\nu} \bm{s}(n)\cdot \bm{s}(n+{\hat\nu}).
\een
The partition function $Z$ is given by
\ben
\label{eq:partitionfunction}
Z=\int{\cal D}[\bm{s}]e^{-S},
\een
where ${\cal D}[\bm{s}]$ is the O(3) Haar measure, whose expression is given later.

\subsection{Tensor network representation of the model}
\label{subsec:tn-rep}

The vector $\bm{s}(n)$ in the model can be expressed as
\begin{equation}
	\label{U:representation}
	\begin{array}{c}
		 \bm{s}^T(\Omega) = (\cos\theta, \sin\theta \cos\phi, \sin\theta \sin\phi) \\
		 \Omega=(\theta,\phi) \quad, ~\theta \in (0,\pi],~\phi \in (0,2\pi]. 
	\end{array}
\end{equation}
The partition function and its measure are written as
\begin{align}
	\label{eq:partitionfunction2}
	Z &= \int {\cal D}\Omega \prod_{n,\nu} e^{\beta  \sum_{i=1}^{3}s_i(\Omega_n) s_i(\Omega_{n+\hat{\nu}})}, \\
	{\cal D}\Omega &= \prod_{p=1}^{V} \frac{1}{4\pi} \sin(\theta_p) d\theta_p d\phi_p~.
\end{align}
We discretize the integration (\ref{eq:partitionfunction2}) with the Gauss-Legendre quadrature~\cite{Kuramashi:2019cgs,Akiyama:2020ntf} after changing the integration variables:
\ben
-1 \le \alpha&=&\frac{1}{\pi}\left(2\theta-\pi \right)\le 1, \\
-1 \le \beta&=&\frac{1}{\pi}\left(\phi-\pi \right)\le 1. 
\een
We obtain
\begin{equation}
	Z = \sum_{ \{\Omega_1\},\cdots,\{\Omega_V\}} \left( \prod_{n=1}^{V} \frac{\pi}{8}  \sin(\theta(\alpha_{a_n})) w_{a_n} w_{b_n} \right) \prod_{\nu} M_{\Omega_n,\Omega_{n+\hat{\nu}}}
\end{equation}
with $\Omega_n=(\theta(\alpha_{a_n}),\phi(\beta_{b_n}))\equiv (a_n,b_n)$, where $\alpha_{a_n}$ and $\beta_{b_n}$ are $a$- and $b$-th roots of the $K$-th Legendre polynomial $P_{K}(s)$ on the site $n$, respectively. $\sum_{ \{\Omega_n\}}$ denotes $\sum_{a_n=1}^{K}\sum_{b_n=1}^{K}$.
$M$ is a 4-legs tensor defined by
\begin{equation}
	M_{a_n,b_n,a_{n+\hat{\nu}}, b_{n+\hat{\nu}}}  =\exp\left\{ \beta \sum_{i=1}^{3}s_i(a_n,b_n) s_i(a_{n+\hat{\nu}}, b_{n+\hat{\nu}}) \right\}~.
\end{equation} 
The weight factor $w$ of the Gauss-Legendre quadrature is defined as
\begin{equation}
	w_{a_n} = \frac{2(1-{\alpha_{a_n}}^2)}{K^2P^2_{K-1}({\alpha_{a_n}})},\quad
	w_{b_n} = \frac{2(1-{\beta_{b_n}}^2)}{K^2P^2_{K-1}({\beta_{b_n}})}.
\end{equation}
After performing the singular value decomposition (SVD) on $M$:
\begin{equation}
	M_{a_n,b_n,a_{n+\hat{\nu}}, b_{n+\hat{\nu}}} \simeq \sum_{i_n=1}^{D_\text{cut}} U_{a_n,b_n, i_n} (\nu) \sigma_{i_n}(\nu) V^\dagger_{i_n,a_{n+\hat{\nu}}, b_{n+\hat{\nu}}} (\nu),
\end{equation}
where $U$ and $V$ denotes unitary matrices and $\sigma$ is a diagonal matrix with the singular values of $M$ in the descending order.
We can obtain the tensor network representation of the O(3) NLSM on the site $n\in\Lambda_{1+1}$
\begin{align}
	T_{x_n, x'_n, y_n, y'_n} &= \frac{\pi}{8} \sqrt{\sigma_{x_n}(1) \sigma_{x'_n}(1) \sigma_{y_n}(2) \sigma_{y'_n}(2) } \sum_{a_n, b_n} w_{a_n} w_{b_n} \nonumber \\
	&\quad \times V^\dagger_{x_n,a_n,b_n} (1) U_{a_n,b_n, x'_n} (1) V^\dagger_{y_n,a_n,b_n} (2) U_{a_n,b_n, y'_n} (2) , 
\end{align} 
where $D_{\text{cut}}$ is the bond dimension of tensor $T$, which controls the numerical precision in the TRG method. The tensor network representation of partition function is given by
\begin{equation}
  Z \simeq \sum_{x_0 x'_0 y_0 y'_0 \cdots} \prod_{n \in \Lambda_{1+1}} T_{x_n x'_n y_n y'_n} = \Tr \left[T \cdots T\right]~.
  \label{eq:Z_TN}
\end{equation}
We employ the higher order tensor renormalization group (HOTRG) algorithm~\cite{PhysRevB.86.045139} to evaluate $Z$.

\subsection{Calculation of entanglement and R\'enyi entropies}
\label{subsec:entropy}

\begin{figure}[H]
	\begin{minipage}[b]{0.3\hsize}
 		\centering
 		\includegraphics[width=0.8\hsize]{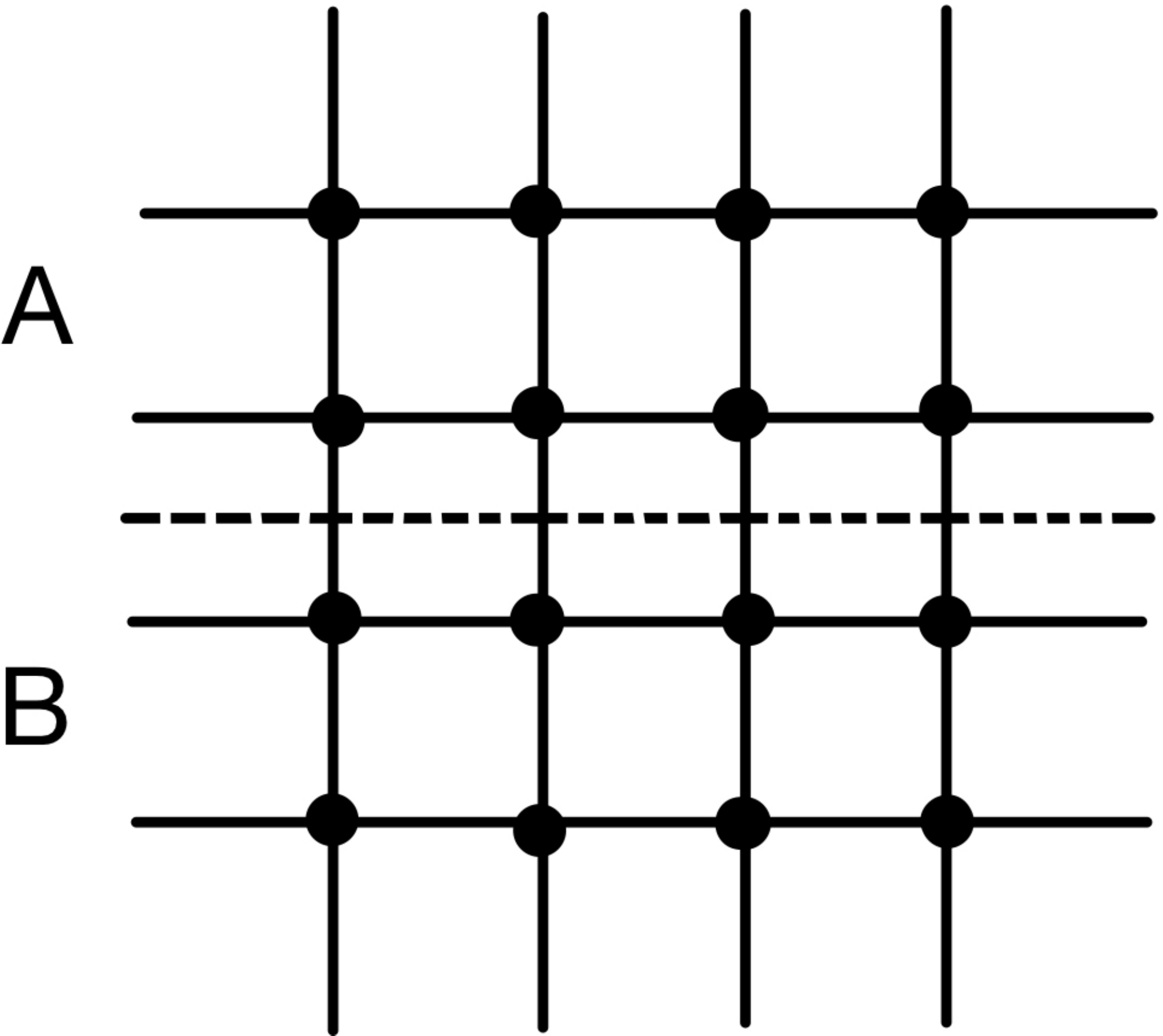}
 		\subcaption{Divide the system to two subsystems A and B.}
 		\label{fig:rhoA1}
	\end{minipage}
	\begin{minipage}[b]{0.3\hsize}
		\centering
		\includegraphics[width=0.85\hsize]{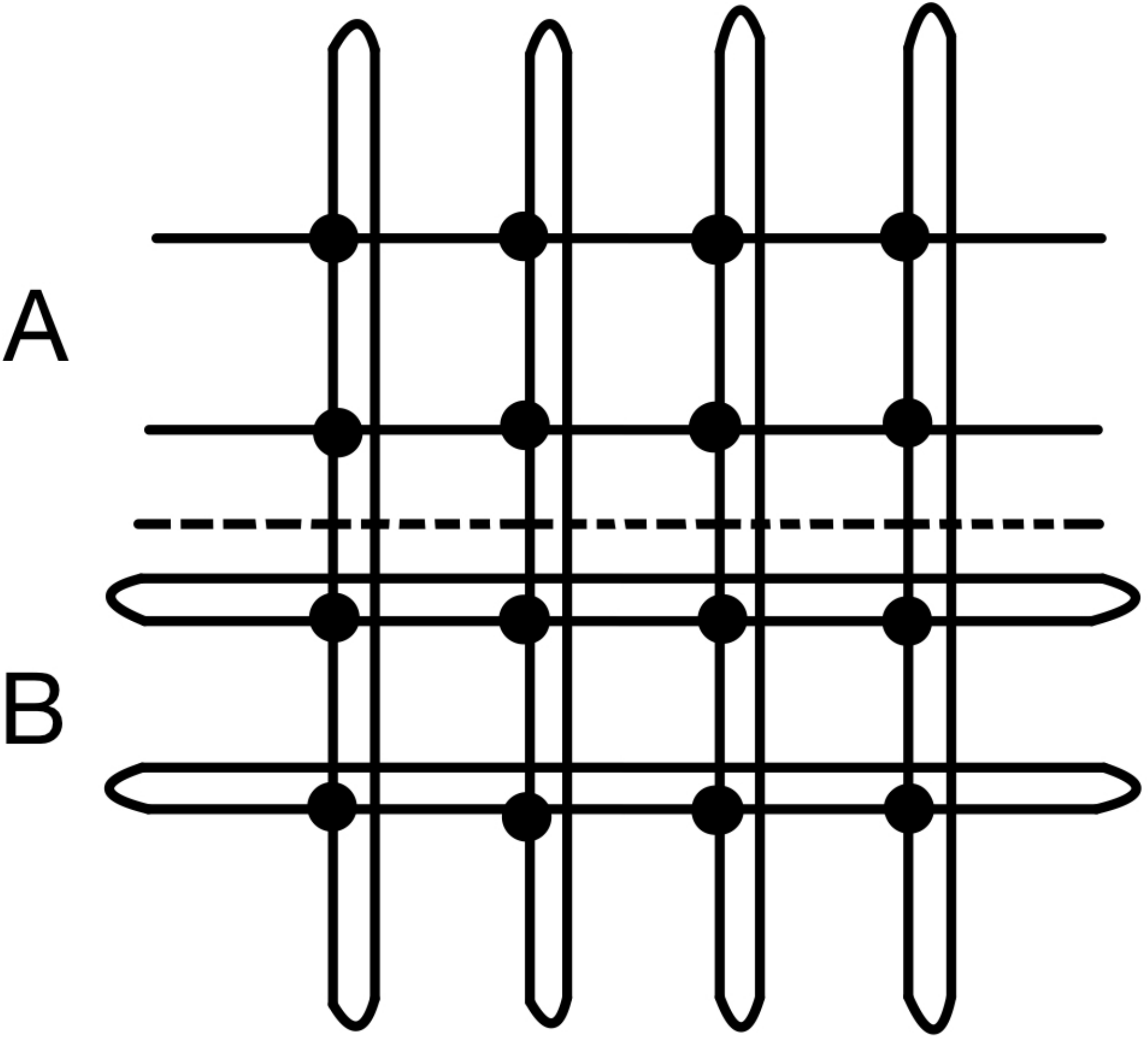}
		\subcaption{Trace all legs for subsystem B.}
		\label{fig:rhoA2}
	\end{minipage}
	\begin{minipage}[b]{0.3\hsize}
		\centering
		\includegraphics[width=0.5\hsize]{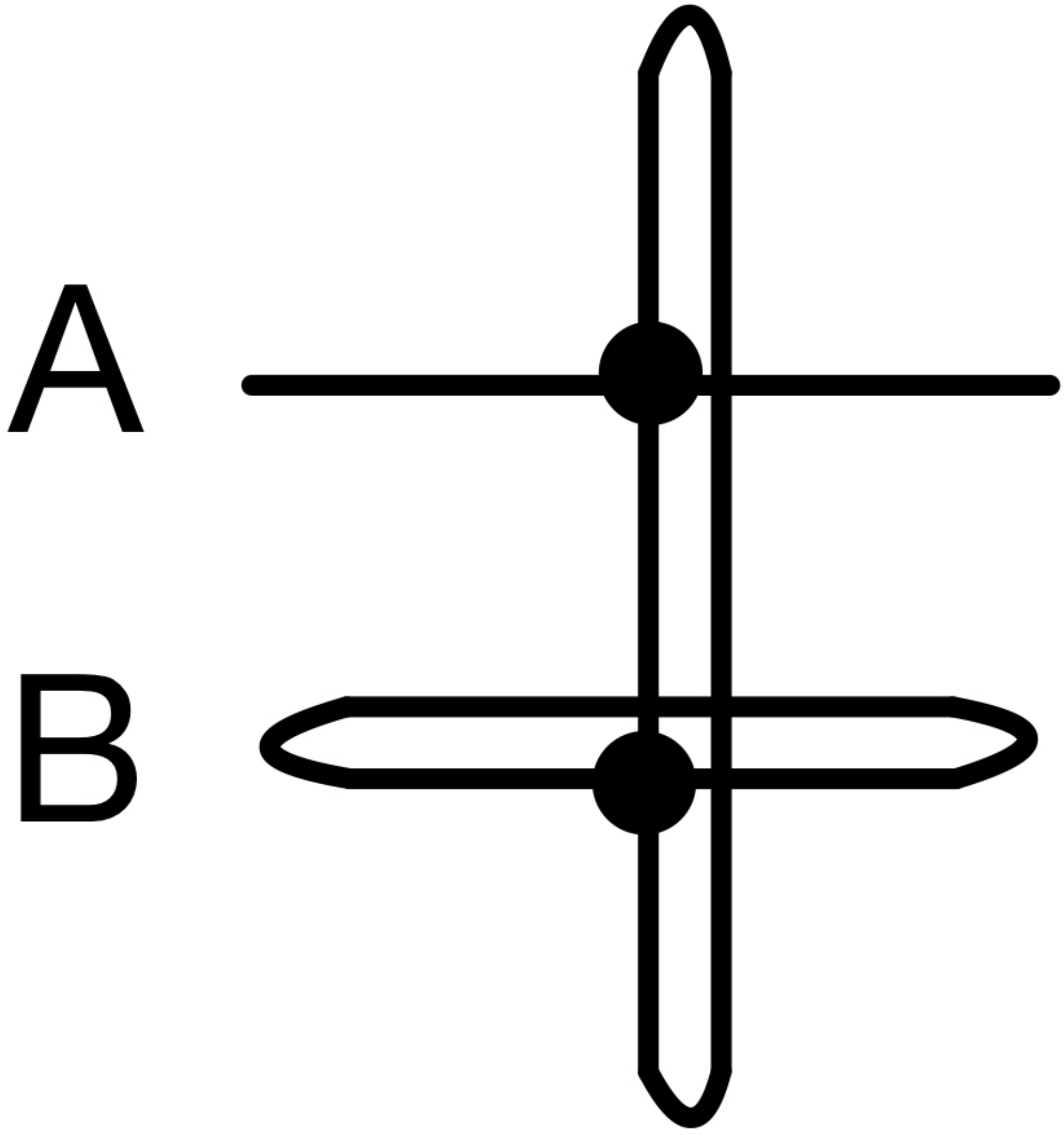}
		\subcaption{Figure~\ref{fig:rhoA2} is coarse-grained to this with HOTRG.}
		\label{fig:rhoA3}
	\end{minipage}
	\caption{Calculation of entanglement entropy.}
	\label{fig:ee_cal}
 \end{figure}

\begin{figure}[H]
	\begin{tabular}{c}
	\begin{minipage}[b]{0.5\vsize}
		\centering
		\includegraphics[width=0.6\vsize]{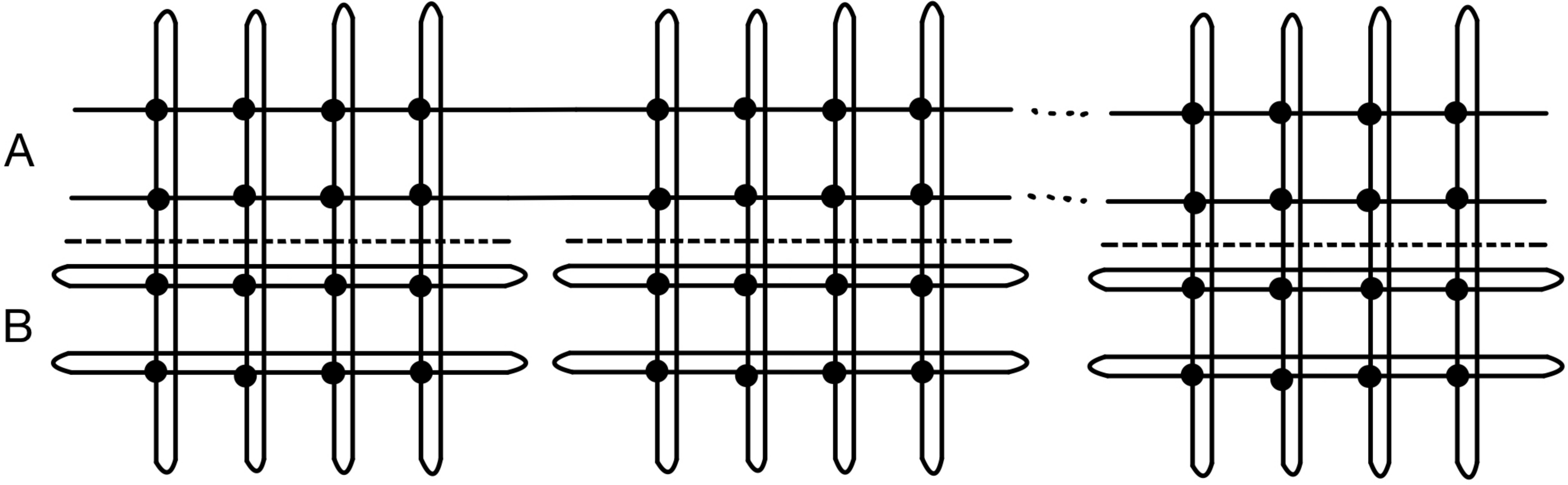}
		\subcaption{Graphical representation of $n$th-order R\'enyi entropy $\rho_A^{(n)}$.}
		\label{fig:renyi1}
	\end{minipage}\\
	\begin{minipage}[b]{0.5\vsize}
		\centering
		\includegraphics[width=0.25\vsize]{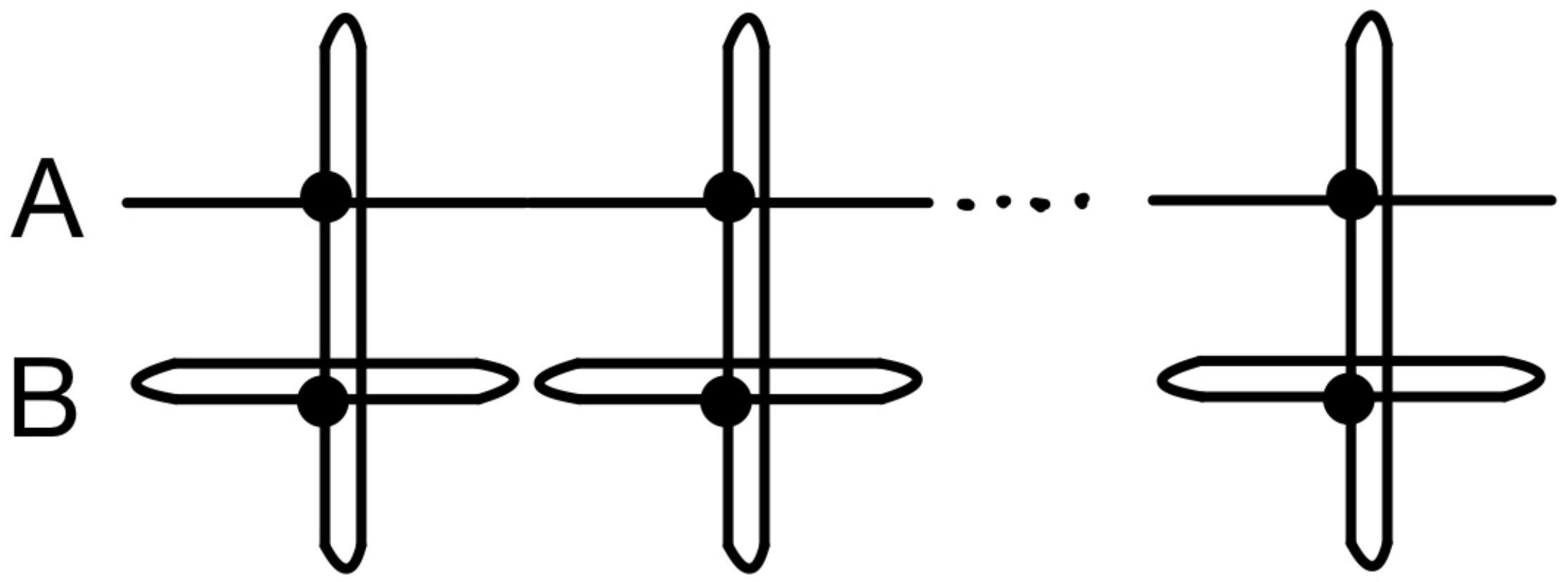}
		\subcaption{Figure~\ref{fig:renyi1} is coarse-grained to this with HOTRG.}
		\label{fig:renyi2}
	\end{minipage}
	\end{tabular}
	\caption{Calculation of $n$th-order R\'enyi entropy.}
	\label{fig:re_cal}
\end{figure}

Figure~\ref{fig:ee_cal} illustrates the calculation procedure of the entanglement entropy. We divide the system to two subsystems A and B, both of which have the same lattice size with $L\times N_t$. The density matrix of subsystem A is defined by $\rho_A={1\over Z}\Tr_B [T\cdots T]$, where $\Tr_B$ denotes the trace restricted to the subsystem B. We use HOTRG to approximate the density matrix of subsystem A, in which $\rho_A\simeq {1\over Z}\Tr_B T_{xx'y_By'_B}T_{x'xy_Ay'_A} = M_{y_A,y'_A}$. The entanglement entropy is obtained by
\ben
S_A = -\Tr_A \rho_A \log(\rho_A).
\een
Figure \ref{fig:re_cal} depicts the calculation procedure of the $n$th-order R\'enyi entropy defined by
\ben
S_A^{(n)} = \frac{\ln \Tr_A \rho_A^n}{1-n},
\een
where $\rho_A^n$ can be calculated by just computing the $n$th matrix power of $\rho_A$.

\section{Numerical results} 
\label{sec:results}


The density matrix $\rho_A$ is evaluated using HOTRG with the bond dimension $D_{\rm cut}\in [10,130]$.
Note that the correlation length $\xi$ in this model was precisely measured over the range of $1.4\le \beta\le 1.9$ with the interval of $\Delta \beta=0.1$ in Ref.~\cite{Wolff:1989hv}. We list the values of $\xi$ in Table~\ref{tab:s} for later convenience. In order to keep the condition $a\ll \xi \ll L$, our results are restricted to $1.4\le \beta \le 1.7$~\footnote{This is an intermediate region from the strong coupling to the weak one. See Fig.~8 in Ref.~\cite{Luo:2022eje}.} in the following.

\begin{figure}[H]
	\centering
	\includegraphics[width=0.75\hsize]{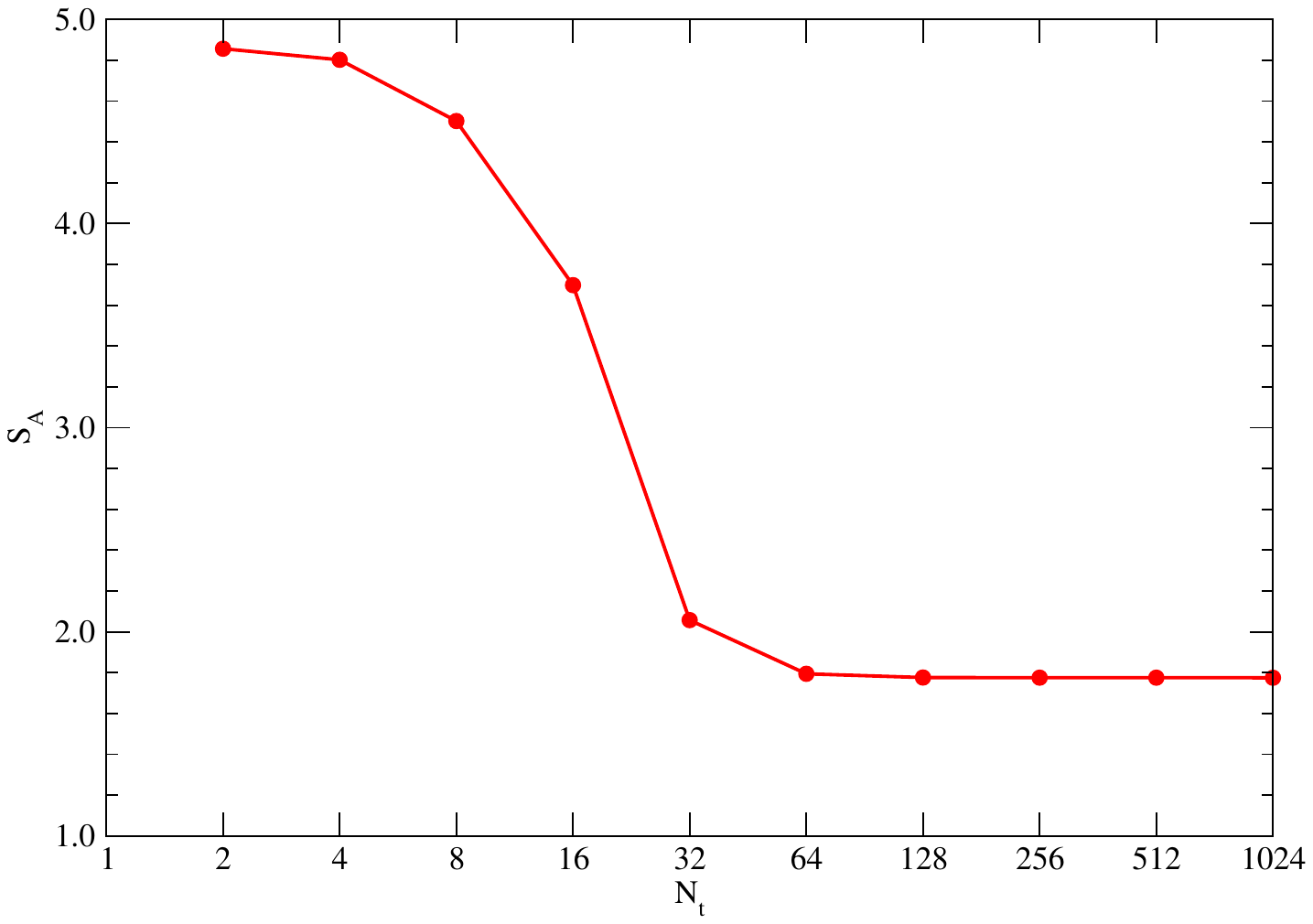}
	\caption{$N_t$ dependence of entanglement entropy at $\beta=1.5$. The bond dimension is $D_{\rm cut}=130$. }
  	\label{fig:sa_nt}
\end{figure}

\begin{figure}[H]
	\centering
	\includegraphics[width=0.75\hsize]{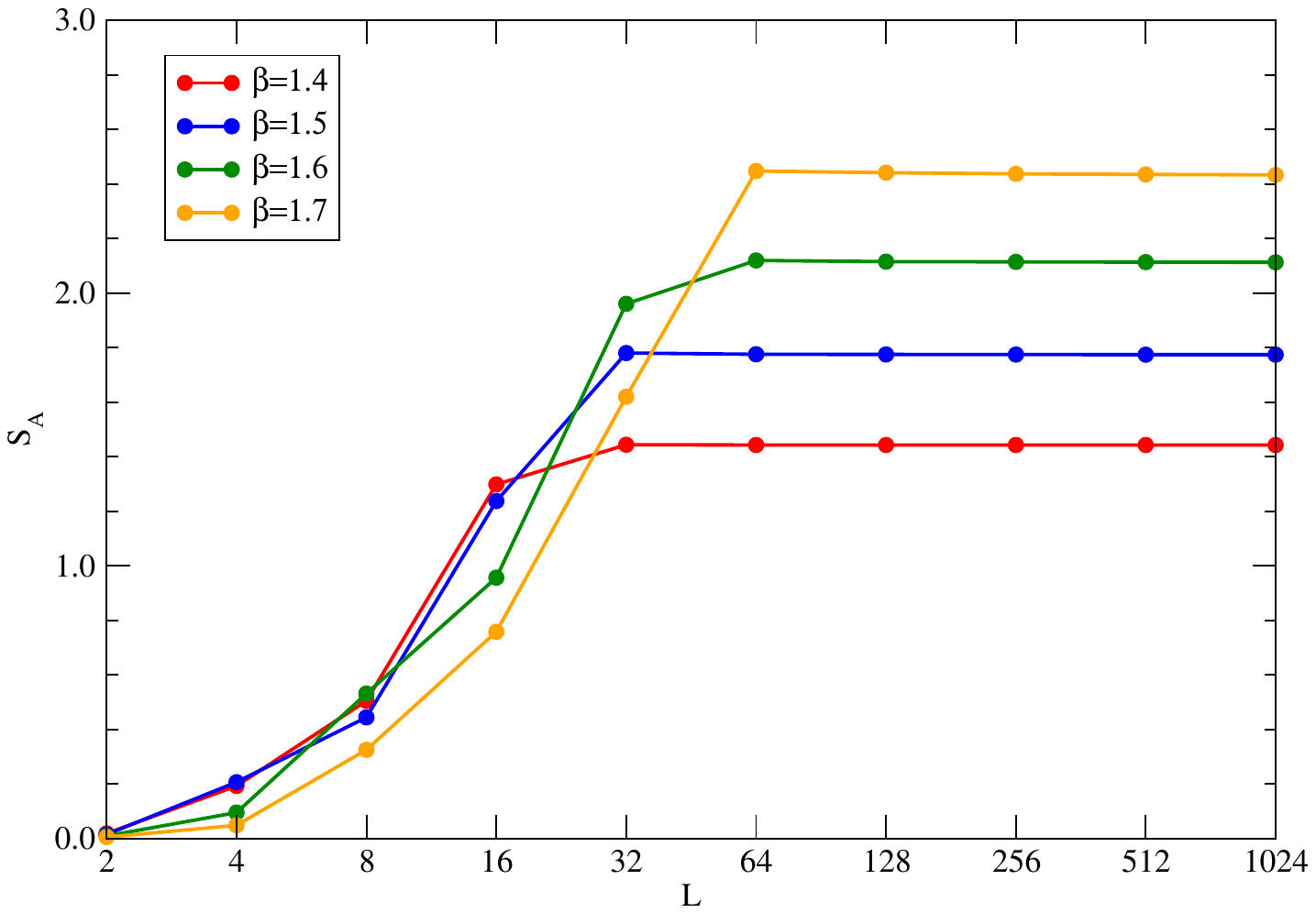}
	\caption{$L$ dependence of entanglement entropy with $N_t=1024$ at $\beta=1.4$, 1.5, 1.6 and 1.7. The bond dimension is $D_{\rm cut}=130$.}
  	\label{fig:sa_beta}
\end{figure}

\begin{figure}[H]
	\centering
	\includegraphics[width=0.75\hsize]{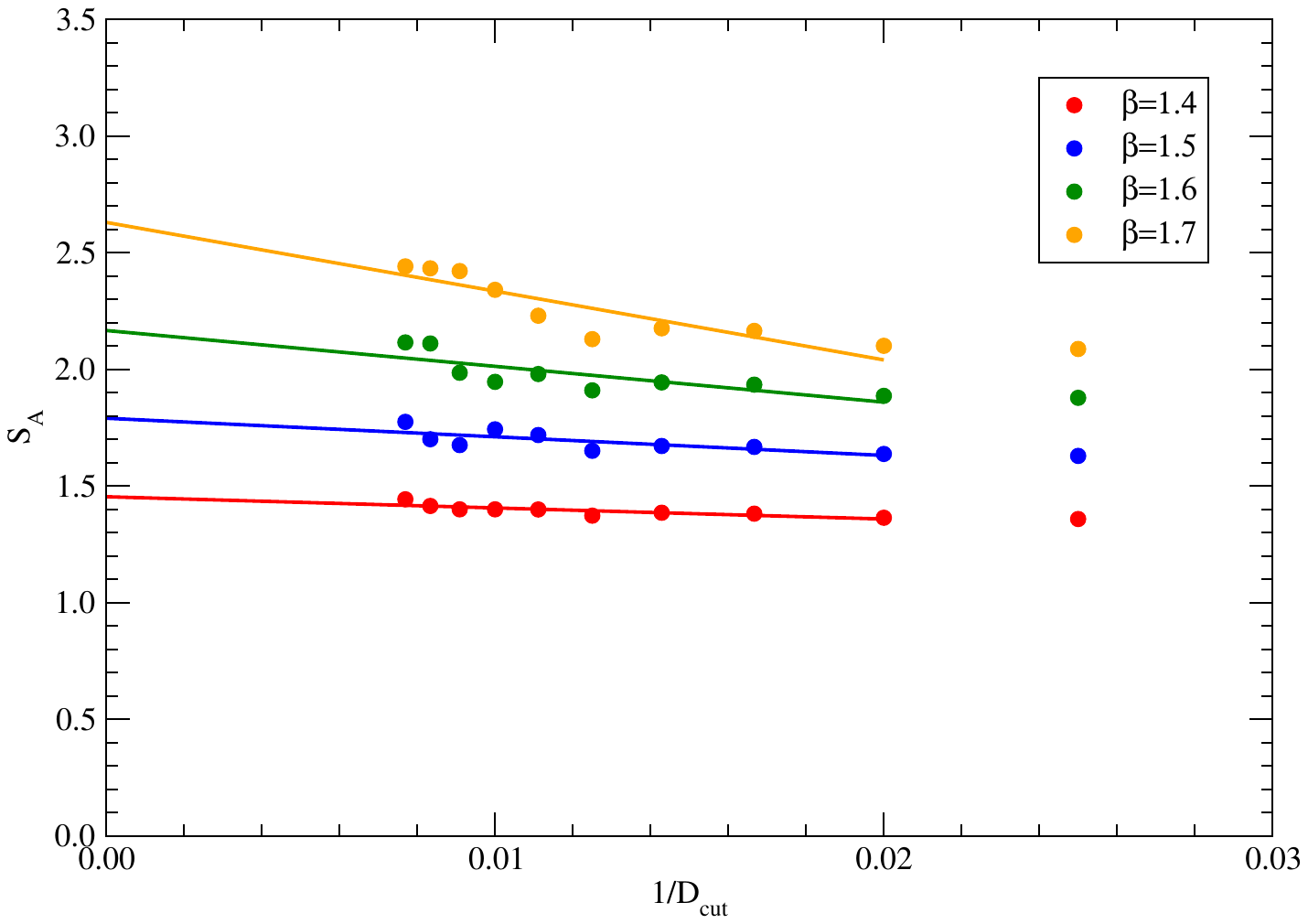}
	\caption{$1/D_{\rm cut}$ dependence of $S_A(L=128)$ with $N_t=1024$ at $\beta=1.4$, 1.5, 1.6 and 1.7. Solid lines denote linear extrapolation.}
  	\label{fig:sa_dinv}
\end{figure}

Figure~\ref{fig:sa_nt} shows the $N_t$ dependence of the entanglement entropy $S_A(L)$ at $\beta=1.5$ with $L=128$, where the correlation length is expected to be $\xi\sim 11$~\cite{Wolff:1989hv}. The degeneracy of the results for $S_A(L)$ with $N_t= 256$, 512 and 1024 indicates the convergence of $S_A(L)$ in terms of $N_t$ so that  $N_t=1024$ is large enough to be regarded as the zero temperature limit. In Fig.~\ref{fig:sa_beta} we plot $S_A(L)$ with $N_t=1024$ at $\beta=1.4$, 1.5, 1.6 and 1.7. The entanglement entropy shows plateau behavior once the interval $L$ goes beyond the correlation length. This is an expected behavior under the condition of $\xi\ll L$~\cite{Calabrese:2004eu}. As $\xi$ increases for larger $\beta$, the plateau of $S_A(L)$ starts at larger $L$ and its value is increased according to the theoretical expectation of $S_A(L)\sim \frac{c}{3}\ln \xi$~\cite{Calabrese:2004eu}. In Fig.~\ref{fig:sa_dinv} we plot $S_A(L=128)$ at $\beta=1.4$, 1.5, 1.6 and 1.7 as a function of $1/D_{\rm cut}$. The data of $S_A(L=128)$ shows increasing trend, while slightly fluctuating, for vanishing $1/D_{\rm cut}$. This kind of fluctuation is commonly observed in the TRG method. See, e.g., Fig. 11 in Ref.~\cite{Ueda_2014} for the Ising model with the HOTRG algorithm. The solid lines express the linear extrapolation of $S_A(L=128)$ at $1/D_{\rm cut}\le 0.02$ to obtain the value at $D_{\rm cut}\rightarrow \infty$, which are listed in Table~\ref{tab:s}.

\begin{figure}[H]
	\centering
	\includegraphics[width=0.75\hsize]{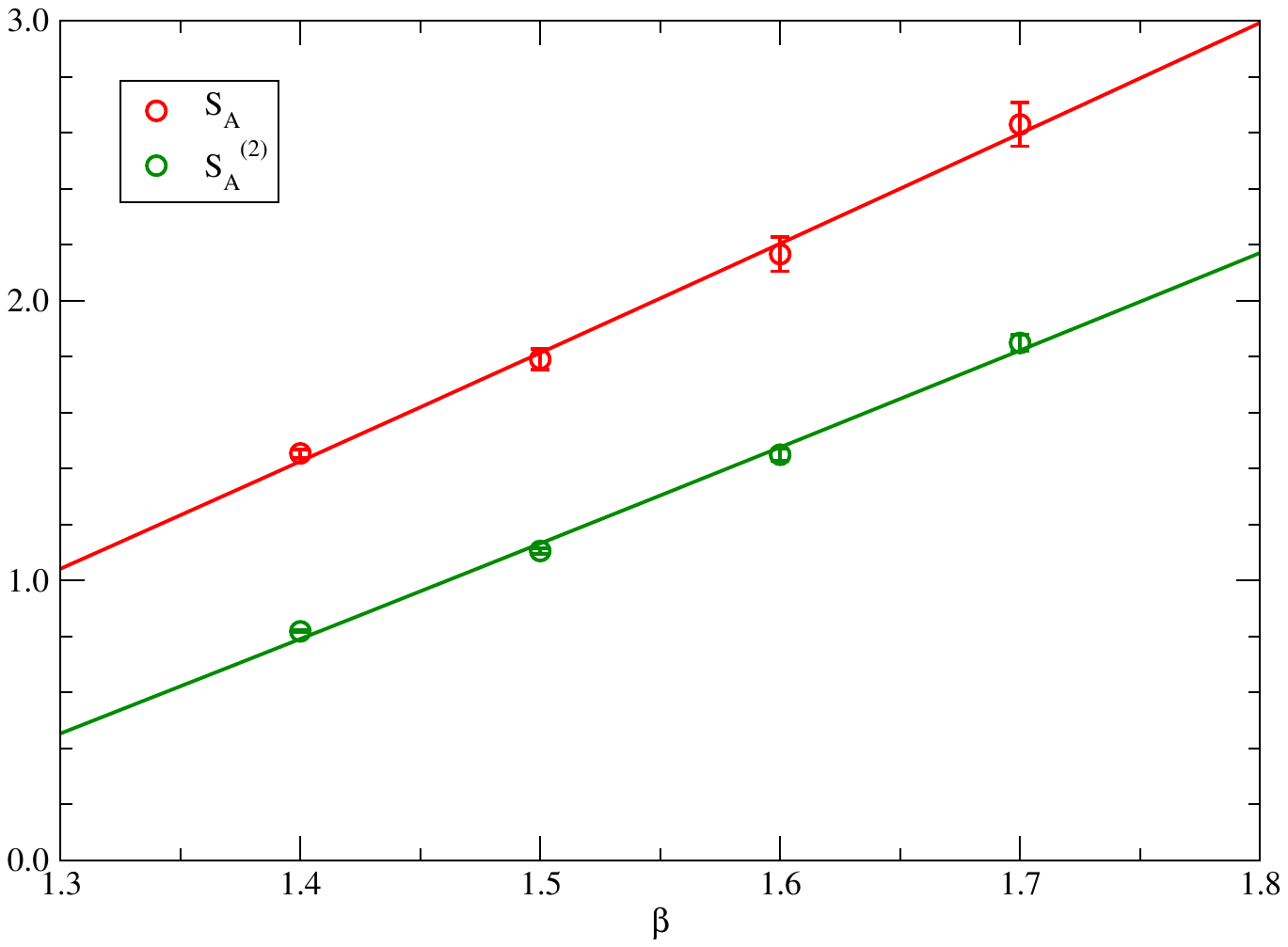}
	\caption{$\beta$ dependence of entanglement and 2nd-order R\'enyi entropies at $L=128$ with $N_t=1024$.}
  	\label{fig:s_beta}
\end{figure}

\begin{figure}[H]
	\centering
	\includegraphics[width=0.75\hsize]{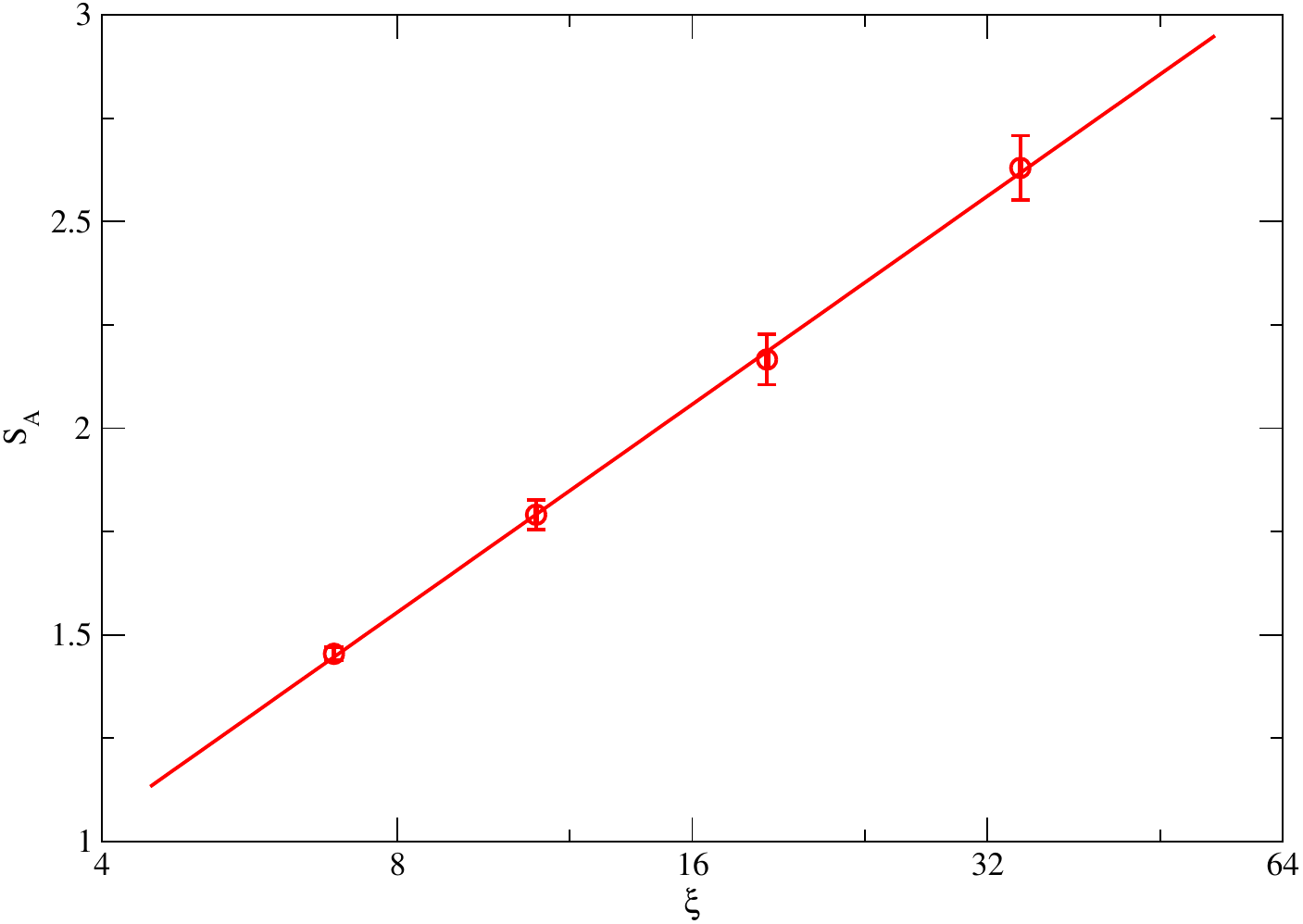}
	\caption{$\xi$ dependence of entanglement entropy at $L=128$ with $N_t=1024$.}
  	\label{fig:s_xi}
\end{figure}

The mass gap $m$ in the (1+1)$d$ O(3) NLSM is expressed as~\cite{Hasenfratz:1990zz}
\ben
m=\frac{8}{e}\Lambda_{\overline{\rm MS}}=64\Lambda_L=\frac{128\pi}{a}\beta\exp\left(-2\pi\beta \right),
\een
where the two-loop expression for the beta function at $\beta\rightarrow \infty$ is used in the last equation. Since the correlation length is inversely proportional to the mass gap the entanglement entropy is rewritten as
\ben
S_A=\frac{c}{3}\left(2\pi\beta-\ln\beta \right)+{\rm const.}
\label{eq:sa_beta}
\een
in terms of the coupling constant $\beta$.
In Fig.~\ref{fig:s_beta} we plot the $\beta$ dependence of $S_A$ at $L=128$ with $N_t=1024$. We determine the central charge $c$ by fitting the data in the range of $1.4\le \beta\le 1.7$ with the function of Eq.~(\ref{eq:sa_beta}), where the condition of $\xi\ll L$ is well satisfied. We obtain the value of $c=1.97(9)$, which is consistent with $c=2.04(14)$ obtained by the MPS method in Ref.~\cite{Bruckmann:2018usp}.  We should also note that a recent study of the central charge for the 2$d$ classical Heisenberg model, which is equivalent to the (1+1)$d$ O(3) NLSM on the lattice, yields $c\sim 2$ with the tensor-network renormalization method~\cite{Ueda_2022}. For an instructive purpose Fig.~\ref{fig:s_xi} shows an alternative plot of $S_A(L=128)$ with $N_t=1024$ as a function of $\xi$ measured in Ref.~\cite{Wolff:1989hv}. This is motivated by a concern that the (1+1)$d$ O(3) NLSM does not have a good asymptotic scaling property below $\beta \sim 2.0$~\cite{Wolff:1989hv,Caracciolo:1994ud}. The use of the fit function $S_A=\frac{c}{3} \ln \xi +{\rm const}.$ gives the central charge $c=2.15(3)$, which is consistent with $c=1.97(9)$ obtained above.

\begin{table}[htb]
	\caption{Results for $S_A(L=128)$ and $S_A^{(n)}(L=128)$ with $N_t=1024$ at $D_{\rm cut}\rightarrow \infty$. Values of correlation length $\xi$ are taken from Ref.~\cite{Wolff:1989hv}.}
	\label{tab:s}
	\begin{center}
	  \begin{tabular}{|c|cccc|}\hline
	    $\beta$ & 1.4 & 1.5 & 1.6 & 1.7 \\ \hline
            $\xi$ & 6.90(1) & 11.09(2) & 19.07(6) & 34.57(7) \\ 
	$S_A$ &  1.45(2)  & 1.79(4) & 2.16(6) & 2.63(8)  \\ 
	$S_A^{(1/2)}$ &  2.39(8) & 2.68(14)  & 3.06(17) & 3.63(17)  \\ 
	$S_A^{(2)}$ &  0.818(4) & 1.105(11)  & 1.449(21) & 1.849(30)  \\ 
	$S_A^{(3)}$ &  0.639(3) & 0.878(8)  & 1.176(15) & 1.535(22)  \\ 
	$S_A^{(4)}$ &  0.569(2) & 0.784(7)  & 1.054(14) & 1.338(19)  \\ 
	$S_A^{(5)}$ &  0.533(2) & 0.735(7)  & 0.989(13) & 1.299(18)  \\ 
	$S_A^{(6)}$ &  0.521(2) & 0.705(7)  & 0.950(12) & 1.247(17)  \\ 
	$S_A^{(7)}$ &  0.498(2) & 0.686(6) & 0.923(12) & 1.213(17) \\ 
	$S_A^{(8)}$ &  0.488(2) & 0.672(6)  & 0.904(17) & 1.188(17)  \\ 
	$S_A^{(9)}$ &  0.480(2) & 0.662(6)  & 0.890(11) & 1.170(16)  \\ 
	$S_A^{(10)}$ & 0.474(2)  & 0.653(6)  & 0.879(11) & 1.155(16)  \\ 
	$S_A^{(11)}$ & 0.469(2)  & 0.647(6)  & 0.870(11) & 1.144(16)  \\ \hline
	\end{tabular}
	\end{center}
\end{table}

\begin{figure}[H]
	\centering
	\includegraphics[width=0.75\hsize]{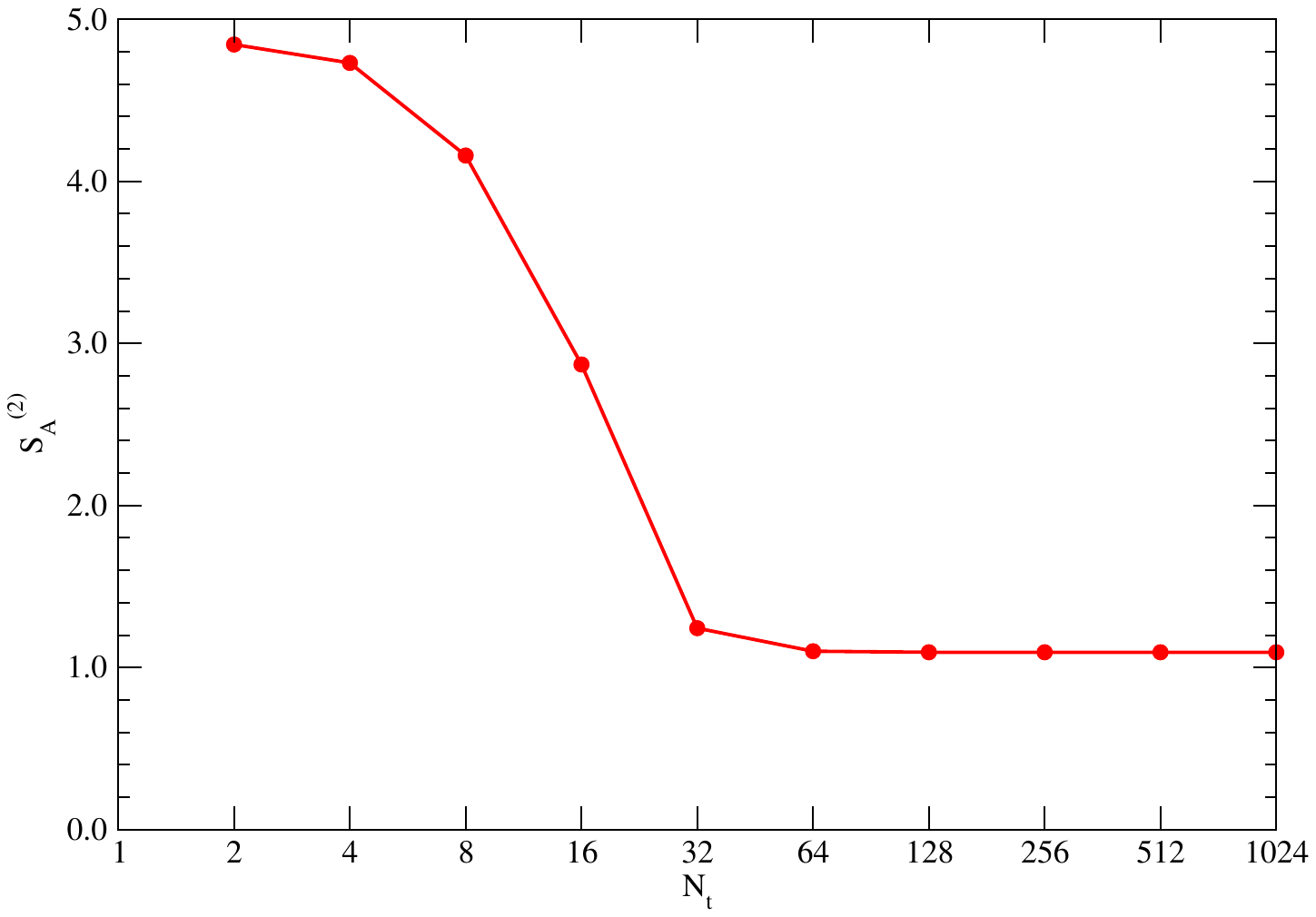}
	\caption{$N_t$ dependence of 2nd-order R\'enyi entropy at $\beta=1.5$. The bond dimension is $D_{\rm cut}=130$.}
  	\label{fig:sa2_nt}
\end{figure}

\begin{figure}[H]
	\centering
	\includegraphics[width=0.75\hsize]{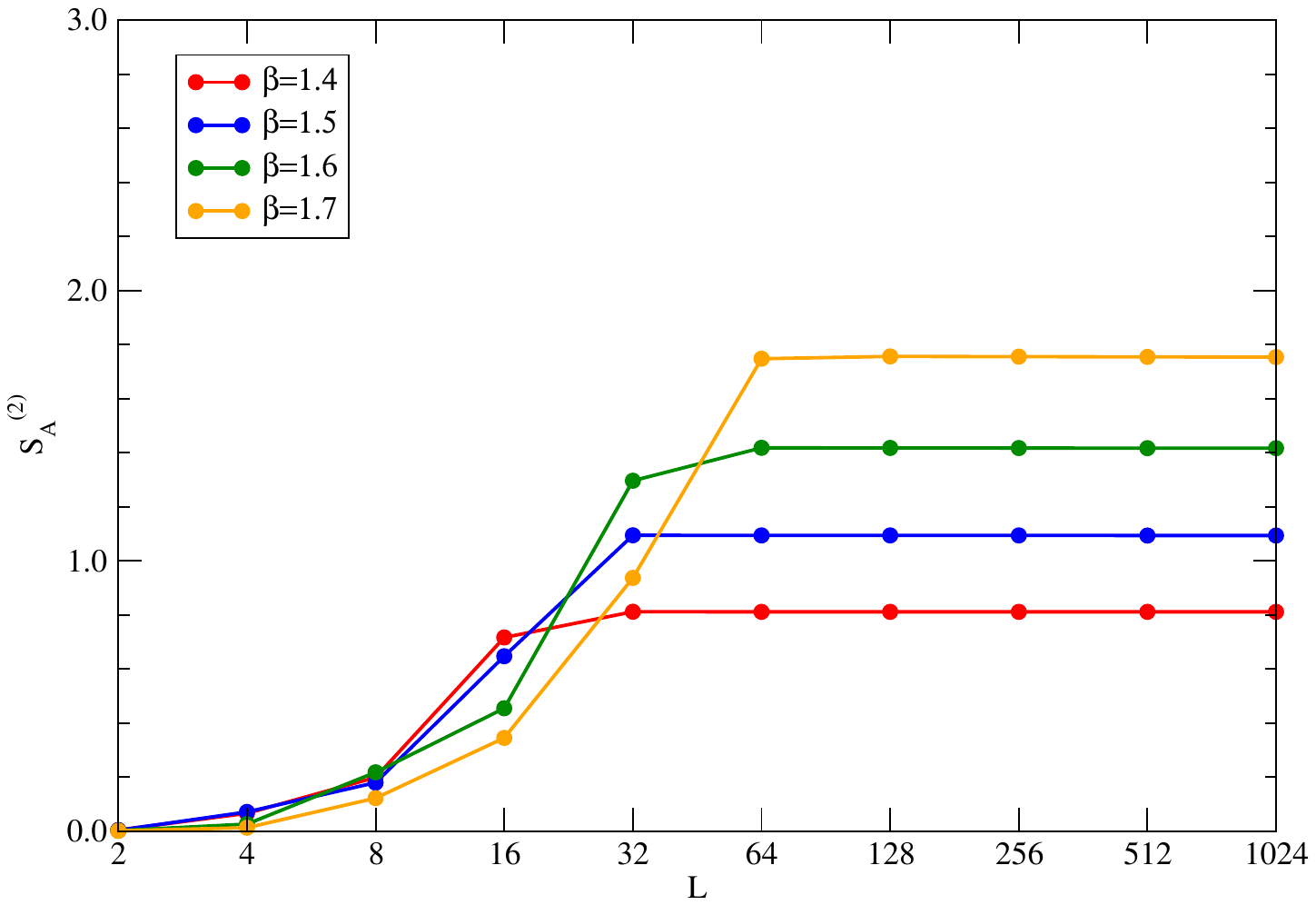}
	\caption{$L$ dependence of 2nd-order R\'enyi entropy with $N_t=1024$ at $\beta=1.4$, 1.5, 1.6 and 1.7. The bond dimension is $D_{\rm cut}=130$.}
  	\label{fig:sa2_beta}
\end{figure}

\begin{figure}[H]
	\centering
	\includegraphics[width=0.75\hsize]{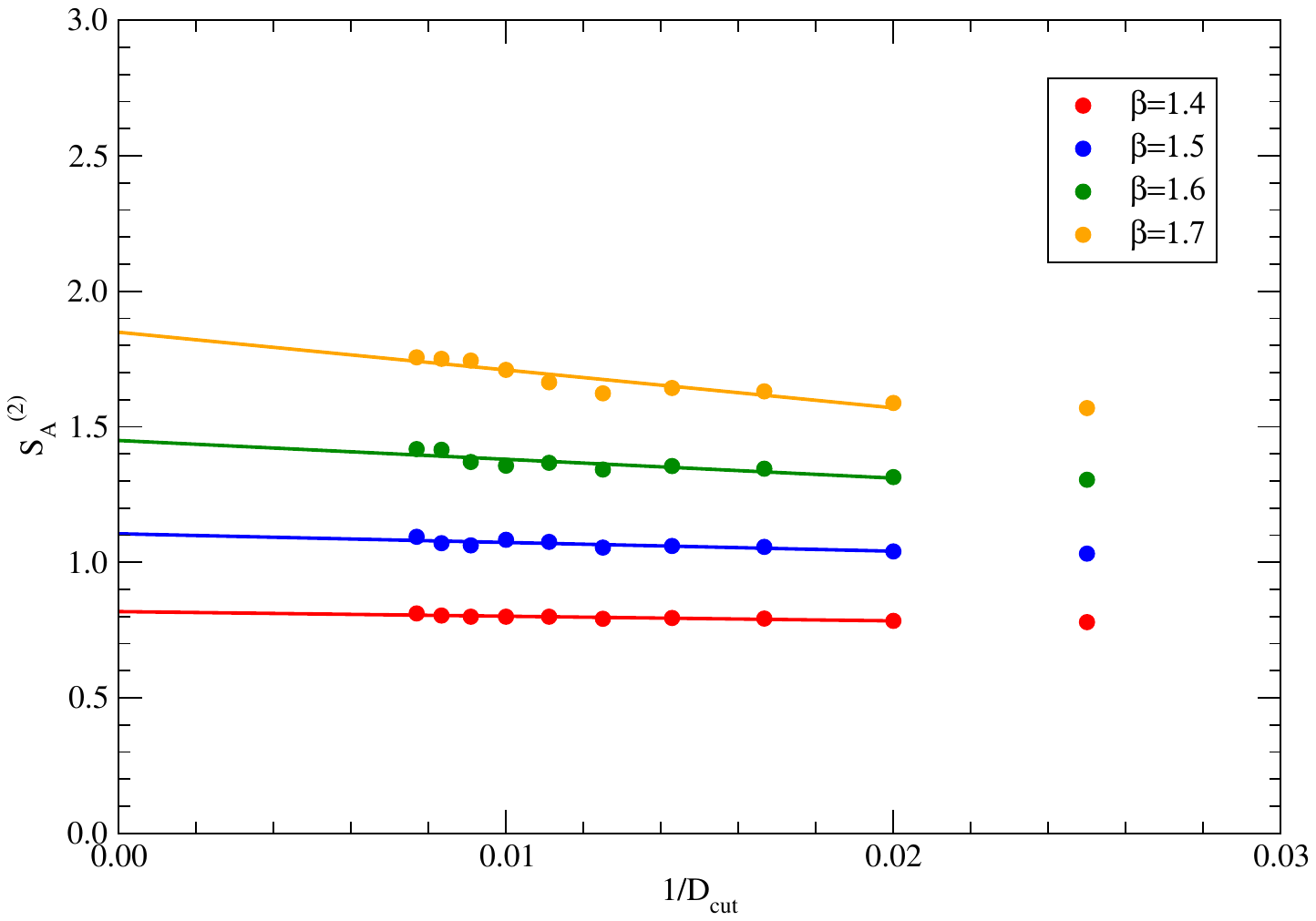}
	\caption{$1/D_{\rm cut}$ dependence of $S_A^{(2)}(L=128)$ with $N_t=1024$ at $\beta=1.4$, 1.5, 1.6 and 1.7. Solid lines denote linear extrapolation.}
  	\label{fig:sa2_dinv}
\end{figure}

Now let us turn to the R\'enyi entropy. In Fig.~\ref{fig:sa2_nt} we plot the $N_t$ dependence of the 2nd-order R\'enyi entropy $S_A^{(2)}(L)$ at $\beta=1.5$ with $L=128$. As in Fig.~\ref{fig:sa_nt} the plateau behavior of $S_A^{(2)}(L)$ is observed in the large $N_t$ region so that $N_t=1024$ is essentially regarded as the zero temperature limit of $S_A^{(2)}(L)$. Figure~\ref{fig:sa2_beta} compares $S_A^{(2)}(L)$ at $\beta=1.4$, 1.5, 1.6 and 1.7 with $N_t=1024$ fixed. Our observation is consistent with the theoretical expectation that $S_A^{(2)}(L)$ should stay constant in the range of $L\gg \xi$ according to $S_A^{(2)}(L)\sim \frac{c}{6}(1+1/n)\ln \xi$~\cite{Calabrese:2004eu}. In Fig.~\ref{fig:sa2_dinv} we show $D_{\rm cut}$ dependence of $S_A^{(2)}(L=128)$ at $\beta=1.4$, 1.5, 1.6 and 1.7. The extrapolated value of $S_A^{(2)}(L=128)$ at $D_{\rm cut}\rightarrow \infty$ is obtained by the linear fit of the data in terms of $1/D_{\rm cut}$ with $1/D_{\rm cut}\le 0.02$. The $\beta$ dependence of $S_A^{(2)}$ at $L=128$ with $N_t=1024$ is plotted in Fig.~\ref{fig:s_beta} together with  $S_A$. We extract the central charge $c$ from the data in $1.4\le \beta\le 1.7$ employing the following fit function with $n=2$: 
\ben
S_A^{(n)}=\frac{c}{6}\left(1+\frac{1}{n}\right)\left(2\pi\beta-\ln\beta \right)+{\rm const.}
\label{eq:san_beta}
\een
The value of $c=2.27(16)$ is slightly larger than that determined from $S_A$.

\begin{figure}[H]
	\centering
	\includegraphics[width=0.75\hsize]{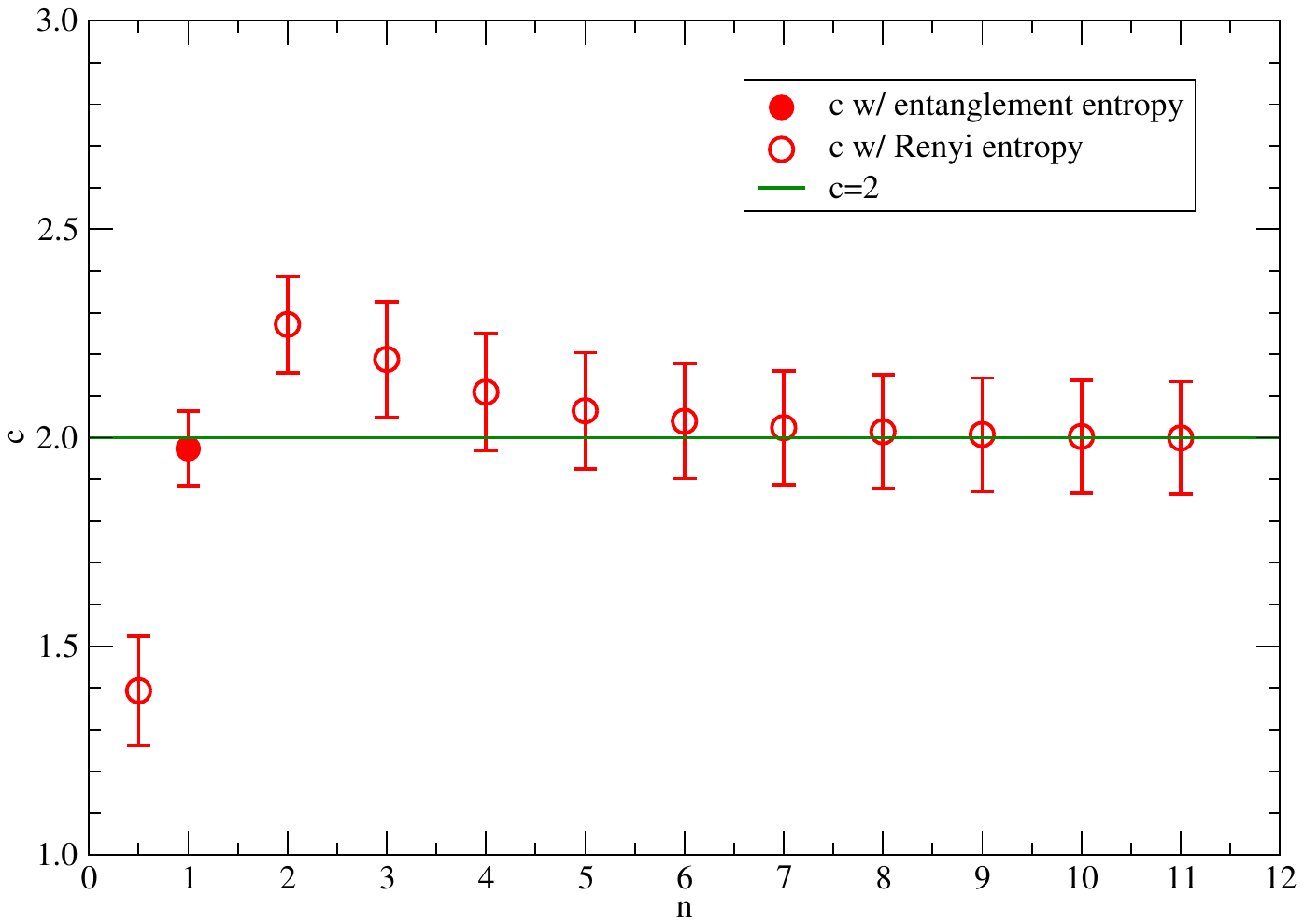}
	\caption{$n$ dependence of central charge $c$ obtained from $n$th-ordr R\'enyi (open) and entanglement (closed) entropies. Solid line denotes $c=2$ to guide your eyes.}
  	\label{fig:cre}
\end{figure}

We repeat the same calculation for other $n$th-order R\'enyi entropy.
The $n$ dependence of the central charge $c$ is plotted in Fig.~\ref{fig:cre},  where the error bar of the central charge originates from the $1/D_{\rm cut}$ extrapolation and the scaling fit with Eq.~(\ref{eq:san_beta}) for the R\'enyi entropy. We observe that the  central value of $c$ seems to converge to $c=2$ as $n$ increases. 
Here we consider the error of the $n$th-order R\'enyi entropy stemming from the errors of the eigenvalues in the density matrix. Suppose ${\bar S}_A^{(n)}$ is the true $n$th-order R\'enyi entropy and  ${\bar \lambda}_j$ denotes the true $j$th eigenvalue in the density matrix $\rho_A$ normalized as $\Tr_A \rho_A=1$:
\ben
{\bar S}_A^{(n)}=\frac{1}{1-n}\ln\Tr_A \rho_A^n=\frac{1}{1-n}\ln \sum_j {\bar \lambda}_j^n,
\een
where we assume the descending order for the eigenvalue ${\bar \lambda}_1>{\bar \lambda}_2>{\bar \lambda}_3,\cdots $. Introducing the error of ${\bar \lambda}_j$, which is expressed as $\delta_j$, the measured R\'enyi entropy may be written as
\ben
S_A^{(n)}=\frac{1}{1-n}\ln \sum_j \left({\bar \lambda}_j+\delta_j \right)^n\simeq \frac{1}{1-n}\ln \sum_j \left({\bar \lambda}_j^n+n\delta_j{\bar \lambda}_j^{n-1}\right).
\een
Focusing on the error of the R\'enyi entropy we find
\ben
\delta S_A^{(n)}&=&\frac{1}{1-n}\frac{1}{\sum_i({\bar \lambda}_i^n)}n\sum_j\delta_j{\bar \lambda}_j^{n-1}
= \frac{n}{1-n}\sum_j\frac{\delta_j}{{\bar \lambda}_j}\frac{1}{\sum_i\left(\frac{{\bar \lambda}_i}{{\bar \lambda}_j}\right)^n},
\een
\ben
\vert \delta S_A^{(n)}\vert &<&\frac{n}{\vert 1-n\vert}\sum_j\frac{\vert \delta_j\vert}{{\bar \lambda}_j}\frac{1}{\sum_i\left(\frac{{\bar \lambda}_i}{{\bar \lambda}_j}\right)^n}
<\frac{n}{\vert 1-n\vert}\sum_j\frac{\vert\delta_j\vert}{{\bar \lambda}_j}\left(\frac{{\bar \lambda}_j}{{\bar \lambda}_1}\right)^n
\rightarrow  \frac{\vert\delta_1\vert}{{\bar \lambda}_1}\;\; (n\rightarrow\infty).
\een
The error of the R\'enyi entropy is bounded by the relative error of the maximum eigenvalue of the density matrix in the large $n$ limit. This may explain the convergence behavior of the central charge {\crr toward $c=2$} observed in Fig.~\ref{fig:cre}.

\begin{figure}[H]
	\centering
	\includegraphics[width=0.75\hsize]{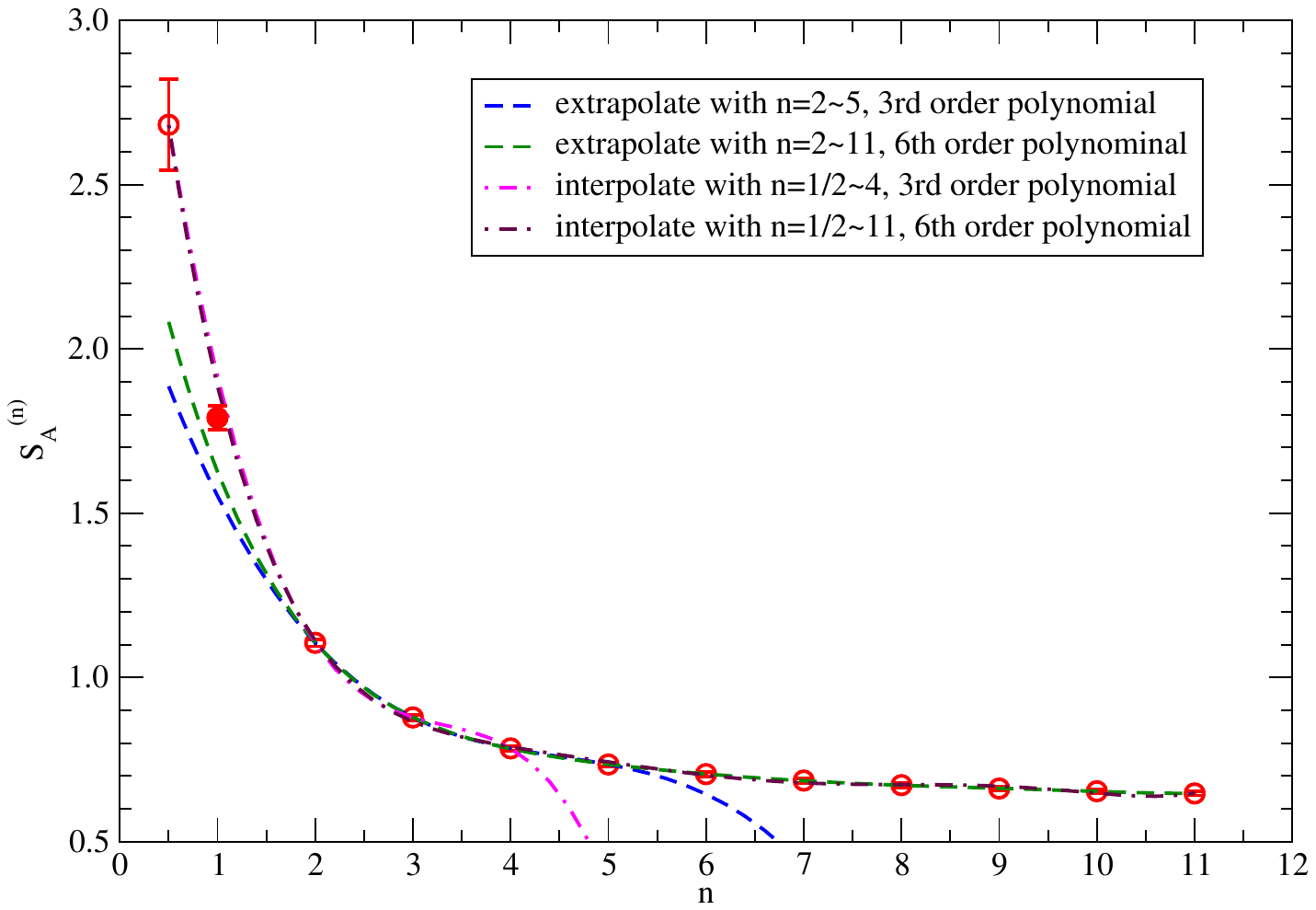}
	\caption{$n$ dependence of $n$th-order R\'enyi entropy with $N_t=1024$ at $\beta=1.5$. Solid symbol at $n=1$ denotes the entanglement entropy. All the results are extrapolated values at $D_{\rm cut}\rightarrow \infty$.}
  	\label{fig:renyi_n}
\end{figure}

In the Monte Carlo approach it is difficult to calculate the entanglement entropy. Actually, previous Monte Carlo studies on the (3+1)$d$ pure SU(N) gauge theories calculate the UV finite observable $\partial_L S^{(2)}_A(L)$ instead of $\partial_L S_A(L)$ assuming $S_A$ is close to $S_A^{(2)}$~\cite{Buividovich:2008kq,Itou:2015cyu,Rabenstein:2018bri}. As observed in Fig.~\ref{fig:s_beta}, the entanglement entropy shows sizable difference from the 2nd-order R\'enyi entropy. This fact implies that the extrapolation of the $n$th-order R\'enyi entropy to $n=1$ might be troublesome. It is worthwhile to check the $n$ dependence of the $n$th-order R\'enyi entropy and investigate how reliably we can extrapolate it to $n=1$. In Fig.~\ref{fig:renyi_n} we plot the $n$-th order R\'enyi entropy as a function of $n$ together with the entanglement entropy at $n=1$. Note that $S_A^{(1/2)}$ is obtained by taking the square root of the density matrix. The dotted blue and green curves represent the fit results of the R\'enyi entropy at $n=2,...,5$ and $n=2,...,11$ employing the third and sixth order polynomial functions, respectively. The extrapolated value to $n=1$ shows sizable deviation from the directly measured entanglement entropy. It seems difficult to obtain the correct value of the entanglement entropy by an extrapolation of the $n$th order R\'enyi entropy at $n\ge 2$. On the other hand, the interpolations of the R\'enyi entropy at $n=1/2,2,...,4$ and $n=1/2,2,...,11$ with the third and sixth order polynomial functions, respectively, which are denoted by the pink and purple curves in Fig.~\ref{fig:renyi_n}, give  consistent results with the entanglement entropy at $n=1$.

\section{Summary and outlook} 
\label{sec:summary}

We have calculated the entanglement and R\'enyi entropies for the (1+1)-dimensional O(3) NLSM under the condition $\xi\ll L$ using the tensor renormalization group method. The central charge obtained from the asymptotic scaling behavior of the entanglement entropy is $c=1.97(9)$, which is consistent with $c=2.04(14)$ previously obtained with the MPS method. We have also investigated the consistency between the entanglement entropy and the R\'enyi entropies. The interpolation using $S_A^{(n)}$ at $n\geq 1/2$ gives a reasonable estimate for $S_A$ at $n=1$, while it is difficult to obtain $S_A$ at $n=1$ from the extrapolation of $S_A^{(n)}$ at $n\geq 2$. As a next step it would be interesting to check the area law in the (2+1)$d$ models.

\begin{acknowledgments}
  Numerical calculation for the present work was carried out with the supercomputer Cygnus under the Multidisciplinary Cooperative Research Program of Center for Computational Sciences, University of Tsukuba.
This work is supported in part by Grants-in-Aid for Scientific Research from the Ministry of Education, Culture, Sports, Science and Technology (MEXT) (No. 20H00148).

\end{acknowledgments}



\bibliographystyle{JHEP}
\bibliography{bib/formulation,bib/algorithm,bib/discrete,bib/grassmann,bib/continuous,bib/gauge,bib/review,bib/for_this_paper}

\end{document}